\documentclass[aps,pre,preprint,groupedaddress]{revtex4-1}
\usepackage{amsmath}
\usepackage{graphicx}
\usepackage{xspace}

\usepackage[thinqspace,cdot]{SIunits}
\usepackage{color}

\def\fig#1{Fig.~\ref{fig:#1}\xspace}
\def\tab#1{Tab.~\ref{tab:#1}\xspace}

\def\avg#1{\left<#1\right>\xspace}

\def\ee{\operatorname{e}}
\def\kb{k_{\mathrm{B}}\xspace}

\def\Myo{Myosin~{II}\xspace}
\def\myo{myosin~{II}\xspace}
\def\myoA{myosin~{IIA}\xspace}
\def\myoB{myosin~{IIB}\xspace}

\def\atp{ATP\xspace}
\def\adp{ADP\xspace}
\def\pho{P$_{\mathrm{i}}$\xspace}

\def\catp{\left[\mathrm{ATP}\right]\xspace}
\def\cadp{\left[\mathrm{ADP}\right]\xspace}
\def\cpho{\left[\mathrm{P_i}\right]\xspace}

\def\fext{F_{\mathrm{ext}}\xspace}
\def\fm{F_{\mathrm{m}}\xspace}
\def\km{k_{\mathrm{m}}\xspace}
\def\fc{F_{\mathrm{c}}\xspace}
\def\fs{F_{\mathrm{s}}\xspace}
\def\epp{E_{\mathrm{pp}}\xspace}
\def\dc{\delta_{\mathrm{c}}\xspace}

\def\nt{N_{\mathrm{t}}\xspace}
\def\nb{N_{\mathrm{b}}\xspace}
\def\zfil{z_{\mathrm{fil}}\xspace}

\def\veff{v_{\mathrm{eff}}\xspace}
\def\vb{v_{\mathrm{b}}\xspace}
\def\vu{v_{\mathrm{u}}\xspace}

\def\fbs{F_{\mathrm{b}}^0\xspace}
\def\fes{F_{\mathrm{eff}}^0\xspace}

\def\seff{\sigma_{\mathrm{eff}}\xspace}
\def\sb{\sigma_{\mathrm{b}}\xspace}
\def\DelG{\Delta G\xspace}
\def\jatp{J_{\mathrm{ATP}}\xspace}

\def\sbm{\sigma^{\mathrm{m}}_{\mathrm{b}}\xspace}
\def\sem{\sigma^{\mathrm{m}}_{\mathrm{eff}}\xspace}

\def\kfe{k_{51}\xspace}
\def\kef{k_{15}\xspace}
\def\kez{k_{12}\xspace}
\def\kze{k_{21}\xspace}
\def\kzd{k_{23}\xspace}
\def\kdz{k_{32}\xspace}
\def\kdv{k_{34}\xspace}
\def\kdf{k_{35}\xspace}
\def\kd{k_{3}\xspace}
\def\kvf{k_{45}\xspace}
\def\kt{k_{\mathrm{T}}\xspace}
\def\katp{k_{\mathrm{T}}\xspace}

\def\pN{\unit{}{\pico\newton}\xspace}
\def\muM{\unit{}{\micro M}\xspace}
\def\miM{\unit{}{\milli M}\xspace}

\def\nm{\unit{}{\nano\meter}\xspace}
\def\nmpers{\unit{}{\nano\meter\per\second}\xspace}

\def\rate#1{\unit{#1}{\second^{-1}}\xspace}
\def\assoc#1{\unit{#1}{\left(\second\usk\micro\mathrm{M}\right)^{-1}}\xspace}
\def\dist#1{\unit{#1}{\nano\meter}\xspace}

\def\force#1{\unit{#1}{\pico\newton}\xspace}
\def\work#1{\unit{#1}{\pico\newton\usk\nano\meter}\xspace}
\def\elast#1{\unit{#1}{\pico\newton\per\nano\meter}\xspace}
\def\mobil#1{\unit{#1}{\nano\meter\per\left(\second\usk\pico\newton\right)}\xspace}

\def\rd{\rho\xspace}

\def\kb{k_{\mathrm{B}}\xspace}
\def\kbt{k_{\mathrm{B}}T\xspace}

\newcommand{\old}[1]{\textcolor{red}{}\xspace}

\begin{document}


\title{Sensitivity of small \myo ensembles from different isoforms\\
to mechanical load and ATP concentration}


\author{Thorsten Erdmann}
\author{Kathrin Bartelheimer}
\author{Ulrich S. Schwarz}
\email{schwarz@thphys.uni-heidelberg.de}
\affiliation{Institute for Theoretical Physics and BioQuant, Heidelberg University, Heidelberg, Germany}


\date{\today}

\begin{abstract}
Based on a detailed crossbridge model for individual \myo motors, we
systematically study the influence of mechanical load and \atp
concentration on small \myo ensembles made from different isoforms.
For skeletal and smooth muscle \myo, which are often used in actomyosin
gels that reconstitute cell contractility, fast forward movement is
restricted to a small region of phase space with low mechanical load and
high \atp concentration, which is also characterized by frequent
ensemble detachment. At high load, these ensembles are stalled or move
backwards, but forward motion can be restored by decreasing \atp
concentration. In contrast, small ensembles of non-muscle \myo isoforms,
which are found in the cytoskeleton of non-muscle cells, are hardly
affected by \atp concentration due to the slow kinetics of the bound states.
For all isoforms, the thermodynamic efficiency of ensemble
movement increases with decreasing \atp concentration,
but this effect is weaker for the non-muscle \myo isoforms.
\end{abstract}


\maketitle



\section{Introduction}

\Myo molecular motors are the main generators
of contractile force in biological systems \cite{b:Howard2001}. As a
non-processive motor, \myo works in groups in order to generate
appreciable levels of force and movement. Although large \myo ensembles
in muscle cells, where a typical ensemble size is $300$ motor heads,
have been extensively studied for decades, only recently
has it become clear that small ensembles of non-muscle isoforms of \myo
are essential for many cellular processes, including cell adhesion,
migration, division and mechanosensing \cite{rv:VicenteEtAl2009,murrell_forcing_2015}. For
example, cellular response to environmental stiffness is abrogated when
\myo is inhibited \cite{a:EnglerEtAl2006}. In the cytoskeleton of non-muscle cells,
\myo is organized in bipolar minifilaments, which are about $300\nm$ in length,
as revealed both by electron \cite{billington_characterization_2013}
and super resolution fluorescence microscopy 
\cite{rv:BeachHammer2015}. In humans, there exist three
non-muscle \myo isoforms. While A and B are both prominent in
determination of cell shape and motility, the role of C is less clear and 
thus we do not discuss it here. The small size of the minifilaments means that cytoskeletal \myo
ensembles contain only $10$-$30$ active motor heads, which limits
their stability because the whole ensemble can stochastically unbind
\cite{a:ErdmannSchwarz2012}.

Outside the cellular context, properties of the main
isoforms of \myo motors (skeletal muscle, smooth muscle, non-muscle A
and B) can be studied in motility assays \cite{rv:VeigelSchmidt2011,
a:WalcottEtAl2012, a:HilbertEtAl2013} and actomyosin gels
\cite{a:SoaresEtAl2011,kohler2011structure,murrell2012f,ideses2013myosin,
linsmeier_disordered_2016}. In the latter case, one often
works with \myo minifilaments from skeletal or smooth muscle, because
they are easier to prepare and to control than those from non-muscle \myo. 
For example, the size of skeletal muscle \myo minifilaments used in a recent actomyosin gel study
has been tuned from $14$ to $144$ \myo molecules using varying salt concentrations \cite{ideses2013myosin}.
While such synthetic skeletal muscle \myo minifilaments seem to have
a very broad size distribution \cite{josephs1966studies},
non-muscle minifilaments from \myo A and B seem to
have a relatively narrow one, close to $30$ \myo molecules 
\cite{niederman1975human,billington_characterization_2013}.
This corresponds to $60$ heads, $30$ for each of the
two ensembles making up the minifilament, of which only a subset is
expected to be active at any moment.
In the cellular context, phosphorylation through regulatory proteins such as 
myosin light chain kinase (MLCK) are required to
make the \myo molecules assembly-competent and to induce motor
activity \cite{somlyo2003ca2+}.

Apart from biochemical modifications of \myo motors due to cellular
signaling, the stochastic dynamics of
small \myo ensembles of a given size is determined mainly by two
physical factors: mechanical load and \atp concentration. From muscle,
it is known that the fraction of bound motors increases under load
\cite{a:PiazzesiEtAl2007}. The underlying molecular mechanism for this
catch bond behavior of \myo is the load dependence of the second phase
of the power stroke, as demonstrated in single molecule experiments
\cite{a:VeigelEtAl2003}. While in muscle this mechanism is used to
stabilize physiological function under load, in non-muscle cells
it is an essential element of the mechanosensitivity of tissue cells
\cite{a:AlbertEtAl2014,a:StamEtAl2015}.

The second physical factor for the dynamics of \myo motors is \atp
concentration, because \atp is required for unbinding from the actin-bound rigor state.
The effect of changes of \atp concentration on the dynamics of \myo
ensembles has been studied before for muscle fibers
\cite{cooke1979contraction}, but not for the small ensembles relevant in
the cytoskeleton, mainly because it is usually assumed that \atp
concentration in tissue cells is constant at a high level around
$1\miM$. However, recently it has been found that \atp
concentration can be much more variable in the cellular context than
formerly appreciated \cite{imamura2009visualization, nakano2011ca2+,
ando2012visualization}. Moreover, reconstitution assays are often
investigated with muscle \myo isoforms at strongly reduced \atp
concentration, but the effect of these differences has not been
systematically studied before.

Here we use a detailed five-state crossbridge
model for single \myo motors to analyze the stochastic dynamics of small \myo ensembles made from
different isoforms as function of both mechanical load and \atp concentration.
Our comprehensive approach combines elements of
earlier models which have used different subsets of mechano-chemical
states \cite{a:Duke1999, a:Duke2000, a:VilfanDuke2003b,
a:WalcottEtAl2012, a:ErdmannSchwarz2012, a:ErdmannEtAl2013,
a:AlbertEtAl2014,a:StamEtAl2015}. By including all relevant states
in one model, we are able to calculate phase diagrams for ensemble performance
as a function of both mechanical load and \atp concentration for all \myo isoforms of interest.
We also discuss the thermodynamic efficiency as a function 
of \atp concentration and find instructive differences between muscle
and non-muscle isoforms.

\begin{figure}[t]
\includegraphics[width=0.8\columnwidth]{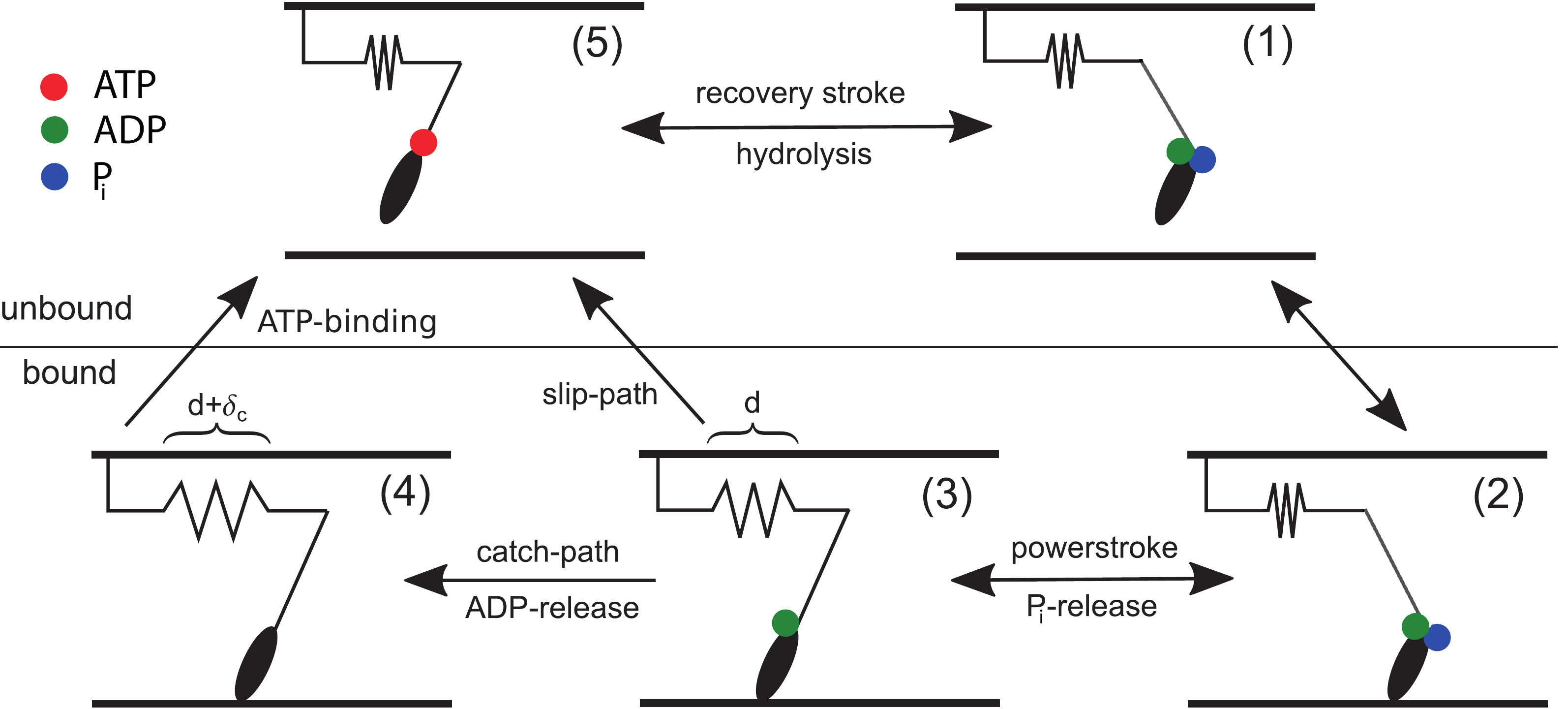}
\caption{Five-state crossbridge model for \myo. Motors stochastically
cycle through five mechano-chemical states with different lever arm
conformations and different combinations of \atp, \adp and \pho bound to
the motor head. Rates for transitions between states depend on
mechanical load and \atp concentration, and are specific 
for the \myo isoform.}
\label{fig:cb_cycle}
\end{figure}

\section{Five-state crossbridge model}

Our crossbridge model for the \myo motor
cycle with five mechano-chemical states is sketched in \fig{cb_cycle}.
In the two states above the line \myo is unbound, while in the three states below
it is bound to the actin filament. The reversible transition
$(5) \to (1)$ with forward rate $\kfe$ and reverse rate $\kef$ is the
recovery stroke. In transition $(1) \to (2)$, \myo motors reversibly
bind to actin with forward rate $\kez$ and reverse rate $\kze$. The
powerstroke $(2) \to (3)$ is driven by a large free energy gain and is
very fast (below milliseconds). The forward rate $\kzd$ is several
orders of magnitude larger than the reverse rate $\kdz$ and here both rates are
assumed to be constant \cite{a:LanSun2005}, although
in practise they might also show some load-dependance. The powerstroke stretches
the elastic neck linker with an effective spring constant $\km$ by a distance $d$.
From state $(3)$, we consider two alternative paths for irreversible
unbinding \cite{a:StamEtAl2015}. The regular motor cycle proceeds from
$(3) \to (4)$ (\emph{catch path}). It requires additional lever arm
movement by $\dc < d$ and is impeded by mechanical load
\cite{a:VeigelEtAl2003}. This load dependence is described by transition
rate $\kdv = \kdv^0 \exp\left( -\fm/\fc \right)$, where $\fm = \km x$ is
the load on a motor with neck linker strain $x$ and $\fc = \kb T/\dc$.
Because the reverse transition requires binding of \adp, which usually is
maintained at very low concentrations, transition $(3) \to (4)$
is considered as irreversible.  Unbinding of \myo from actin in
transition $(4) \to (5)$ requires binding of \atp. This is described by
the transition rate $\kvf = \kt \catp$. Alternatively, motors can unbind
directly from state $(3)$ to $(5)$ along the \emph{slip path} with
transition rate $\kdf = \kdf^0 \exp\left( \fm/\fs \right)$, which
increases with the load $\fm$. The slip path for unbinding
has been demonstrated in single molecule experiments
\cite{a:GuoGuilford2006}. With $\kdv^0 \gg \kdf^0$ and $\fs \gg \fc$,
it is activated only under large load and prevents stalling of the motor cycle.

\begin{table}[h!t]
\begin{center}
\begin{tabular}{l|c|c|c|c|c|c|c}
\hline\hline
Parameter              & Symbol                 & Units      & Skeletal & Smooth & NM IIA & NM IIB & References                                 \\
\hline\hline
Transition rates       & $\kez$                 & $\rate{}$  & $40$     & $6$    & $0.2$   & $0.2$   & \cite{a:VilfanDuke2003b,a:WalcottEtAl2012, a:StamEtAl2015} \\
                       & $\kze$                 & $\rate{}$  & $2$      & $2$    & $0.004$ & $0.004$ & \cite{a:Duke1999,kovacs2003}             \\
							  & $\kzd = \kzd^0\ee^{-\frac{\epp}{2\kbt}}$ & $\rate{}$ & $1.4\cdot 10^6$    & $1.4\cdot 10^6$    & $1.4\cdot10^6$    & $1.4\cdot 10^6$    & \cite{a:VilfanDuke2003b}                        \\
                       & $\kdz = \kdz^0\ee^{+\frac{\epp}{2\kbt}}$ & $\rate{}$ & $7  \cdot 10^{-1}$ & $7  \cdot 10^{-1}$ & $7  \cdot10^{-1}$ & $7  \cdot 10^{-1}$ & \cite{a:VilfanDuke2003b}                        \\
							  & $\kzd^0 = \kdz^0$      & $\rate{}$  & $1000$   & $1000$ & $1000$  & $1000$  & \cite{a:VilfanDuke2003b}                        \\
                       & $\kdv^0/\kd^0$         & $-$        & $0.92$   & $0.92$ & $0.92$  & $0.92$  & \cite{a:GuoGuilford2006,a:StamEtAl2015}                  \\
                       & $\kdf^0/\kd^0$         & $-$        & $0.08$   & $0.08$ & $0.08$  & $0.08$  & \cite{a:GuoGuilford2006,a:StamEtAl2015}                  \\
                       & $\kd^0$                & $\rate{}$  & $350$    & $18$   & $1.71$  & $0.35$  & \cite{a:VilfanDuke2003b,a:WalcottEtAl2012,a:StamEtAl2015}   \\
                       & $\katp$                & $\assoc{}$ & $2$      & $1.2$  & $1.2$   & $1.2$   & \cite{a:WalcottEtAl2012}                       \\
                       & $\kfe$                 & $\rate{}$  & $100$    & $100$  & $100$   & $100$   & \cite{a:WalcottEtAl2012}                       \\ 
							  & $\kef$                 & $\rate{}$  & $10$     & $10$   & $10$    & $10$    & \cite{a:WalcottEtAl2012}                       \\
Powerstroke distances   & $d$                    & $\dist{}$  & $10$     & $10$   & $5.5$   & $5.5$   & \cite{a:WalcottEtAl2012,a:StamEtAl2015}              \\
                       & $\dc$                  & $\dist{}$  & $1.86$   & $2.60$ & $2.5$   & $2.5$   & \cite{a:WalcottEtAl2012,a:StamEtAl2015}              \\
Unbinding forces        & $\fc$                  & $\force{}$ & $2.23$   & $1.59$ & $1.66$  & $1.66$  & \cite{a:WalcottEtAl2012,a:StamEtAl2015}              \\
                       & $\fs$                  & $\force{}$ & $13.91$  & $9.95$ & $10.35$ & $10.35$ & \cite{a:WalcottEtAl2012,a:StamEtAl2015,a:GuoGuilford2006}      \\
Powerstroke bias       & $\epp$                 & $\work{}$  & $-60$    & $-60$  & $-60$   & $-60$   & \cite{a:VilfanDuke2003b}                        \\
Neck linker elasticity & $\km$                  & $\elast{}$ & $0.3$    & $0.3$  & $0.7$   & $0.7$   & \cite{a:WalcottEtAl2012, a:StamEtAl2015}             \\		
Thermal energy         & $\kbt$                 & $\work{}$  & $4.14$   & $4.14$ & $4.14$  & $4.14$  & \cite{a:VilfanDuke2003b}                        \\		
Mobility               & $\eta$                 & $\mobil{}$ & $1000$   & $1000$ & $1000$  & $1000$  & \cite{a:ErdmannEtAl2013}                       \\
\hline
Duty ratio             & $\rho$                 & $-$        & $0.1$ & $0.25$ & $0.1$      & $0.36$  & $-$ \\ 
\hline\hline
\end{tabular}
\caption{Model parameters for different \myo isoforms as extracted from the literature.}
\label{tab:general_rates}
\end{center}
\end{table}

In \tab{general_rates} we list the molecular parameters and transition rates of our model for
four different \myo isoforms as extracted from the literature. Following our earlier work on \myo ensembles
\cite{a:ErdmannSchwarz2012, a:ErdmannEtAl2013, a:AlbertEtAl2014}, the parameters for
skeletal muscle \myo are used as the reference case
which here is compared to results for other \myo isoforms.
Parameters for skeletal and smooth muscle
\myo are taken from Ref.\ \cite{a:WalcottEtAl2012} and for non-muscle \myoA and B
from Ref.\ \cite{a:StamEtAl2015}. Parameters not included in those models are
supplemented from Refs.\ \cite{a:VilfanDuke2003b, a:WalcottEtAl2012, a:GuoGuilford2006}. It should be noted
that literature values for powerstroke distance $d$ and neck linker elasticity $\km$ are usually
effective quantities obtained by fitting procedures and vary
significantly even for the same isoform. For skeletal and smooth muscle
\myo, we use the small value $\elast{0.3}$ given in Ref.\ \cite{a:WalcottEtAl2012}.
For non-muscle \myoA and B, on the other hand, we use the larger value
$\km = \elast{0.7}$ used in Ref.\ \cite{a:StamEtAl2015}. Parameters in
Ref.\ \cite{a:WalcottEtAl2012} are obtained from fits to laser trap experiments and
motility assays for small \myo ensembles so that compliance of the
environment might contribute to the smaller neck linker stiffness.
Parameters in Ref.\ \cite{a:StamEtAl2015} are based on single molecule experiments.
Moreover, the parameters from Ref.\ \cite{a:WalcottEtAl2012} yield larger values for
the single motor duty ratio at vanishing load and large \atp
concentration than observed in muscle.  The single motor duty ratio
$\rd$ describes the probability that a motor is bound to the substrate.
For a two-state model, it would be $k_{\rm on}/(k_{\rm on}+k_{\rm
off})$. Due to the large powerstroke rate $\kzd$, the single motor duty
ratio for vanishing load and large \atp concentration can be estimated
as $\rd \simeq \kez / (\kez + \kd^0)$ as done in the last line 
of \tab{general_rates}.

In a \myo ensemble, $\nt$ individual motors are coupled to the rigid
motor filament via their elastic neck linkers. The state of
an ensemble is characterized by the mechano-chemical states of all
motors and the positions of bound motor heads on the actin filament. For
given external load $\fext$, the position $\zfil$ of the motor filament is
adjusted dynamically by the balance of external load and elastic motor
forces $\fm = \km x$ of all bound motors \cite{a:ErdmannEtAl2013}.
The resulting bound velocity $\vb$ is averaged to give a measure
for how well the ensemble is advancing. Although single motors usually step
only forward, the filament can also move backward if unbound motors
rebind behind the average position of bound motors heads on the substrate.
Moreover, due to the small
ensemble size, it can happen that all motors are unbound at once.
In this case, a different physical process
has to take over to determine how fast the ensemble is moving.
Here we assume that while the filament is unbound, it is pulled
backwards with unbound velocity $\vu = - \eta\fext$,
until a first motor binds through transition $(1) \to (2)$.
Due to this important effect, the resulting effective velocity $\veff$ is
smaller than the bound velocity $\vb$. Here we analyze the dynamics of
\myo ensembles numerically using exact stochastic simulations with the
Gillespie algorithm. For more details on these procedures, we refer to
our earlier work \cite{a:ErdmannSchwarz2012, a:ErdmannEtAl2013, a:AlbertEtAl2014}.

\begin{figure}[t]
\includegraphics[width=0.8\columnwidth]{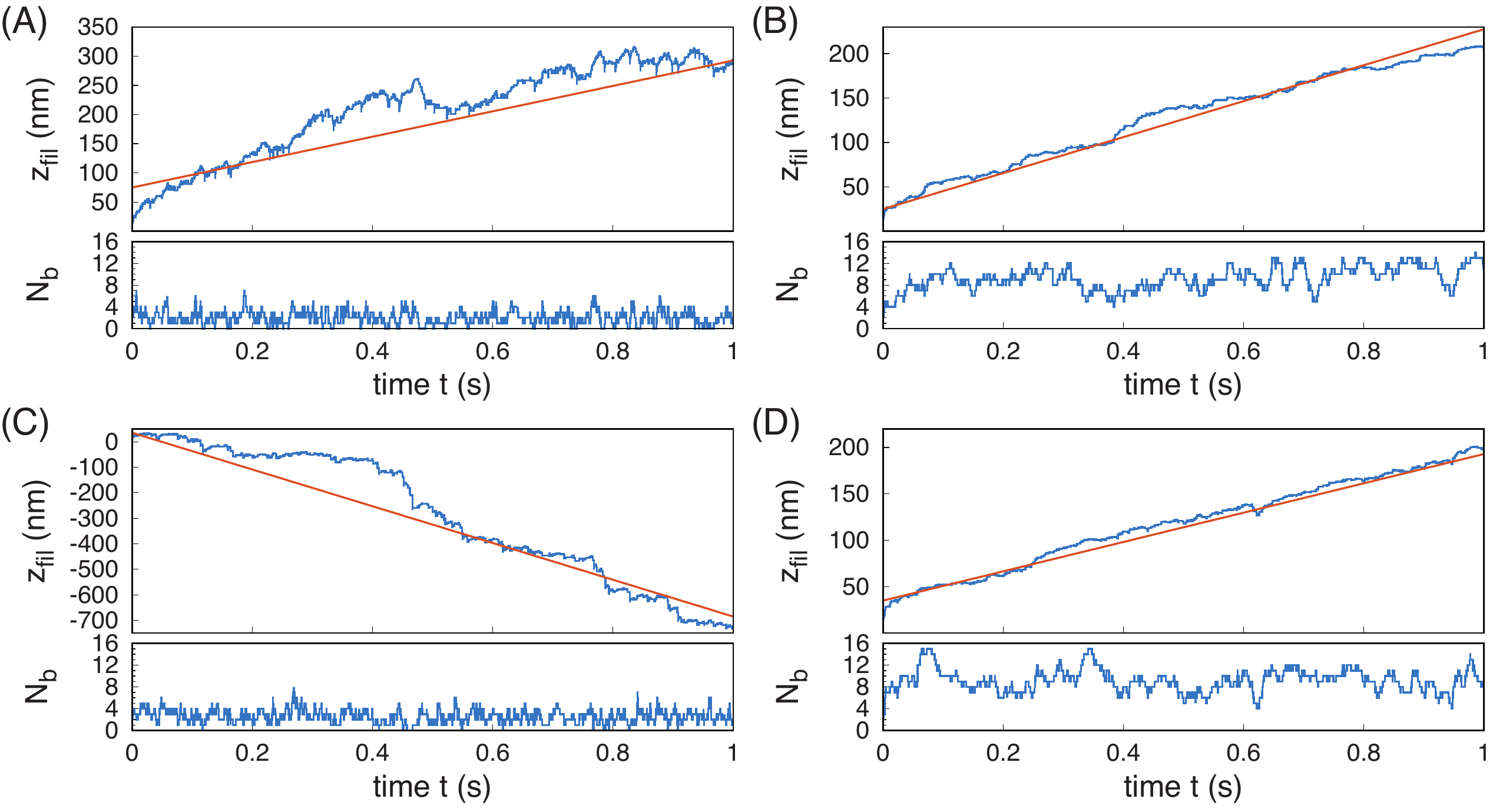}
\caption{Stochastic trajectories for a skeletal muscle \myo ensemble of size $\nt = 16$.
Each trajectory displays the fluctuating number $\nb$ of bound motors
(lower panel) and the position $\zfil$ of the motor filament (upper
panel) as function of time. The straight lines indicate movement
with average effective velocity $\avg{\veff}$. Trajectories are shown
for small and large values of external load and \atp concentration:
$\fext/\nt = 0.2\pN$ in (A,B) (top) and $\fext/\nt = 0.5\pN$ in (C,D)
(bottom); $\catp = 1\miM$ in (A,C) (left) and $\catp = 10 \muM$ in (B,D)
(right).}
\label{fig:trajectories}
\end{figure}


\section{Trajectories}

\fig{trajectories} shows typical stochastic trajectories for a skeletal
\myo minifilament of size $\nt = 16$, which is
a typical value for the number of active \myo motor heads
in the minifilament ensembles used in motility assays and in the cytoskeleton
of non-muscle cells. For each trajectory, lower and upper panels display
the fluctuating number $\nb$ of bound motors and the fluctuating
position $\zfil$ of the motor filament, respectively.  The stochastic
trajectory of $\zfil$ is compared to movement with average effective
velocity $\avg{\veff}$.

In \fig{trajectories}(A), mechanical load is small while \atp
concentration is large and comparable to cellular concentrations.  Due
to the small single motor duty ratio, the average number of bound motors
is small, $\avg{\nb} \simeq 1.9 \ll \nt$, and the ensemble frequently
detaches completely. For the small load, both bound and effective
velocity are positive, although $\avg{\veff} \simeq 220\nmpers$ is
significantly smaller than $\avg{\vb} \simeq 540\nmpers$.
\fig{trajectories}(B) demonstrates the stabilizing effect of decreased
\atp concentration. The average number of bound motors increases to
$\avg{\nb} \simeq 8.8$ and ensemble detachment is no longer observed.
Bound and effective velocity are therefore identical, $\avg{\vb} \simeq
\avg{\veff} \simeq 200\nmpers$, but both are smaller than for large
\atp concentration. \fig{trajectories}(C) demonstrates the stabilizing
effect of increased external load at the same high \atp concentration as
in (A). The average number of bound motors now is $\avg{\nb} \simeq 2.6$
and complete detachment occurs less frequently than in (A). Because the
high load opposes movement, the ensemble now moves backwards with
$\avg{\vb} \simeq -520\nmpers$ and $\avg{\veff} \simeq -720\nmpers$.

\fig{trajectories}(D) demonstrates the effect of reduced \atp
concentration at large external load. Compared to (A), the average number of bound motors
is increased to $\avg{\nb} \simeq 8.8$. As in (B), ensemble detachment
does not occur so that bound and effective velocities are identical. In
contrast to the case of small load, backward movement observed at large
load and large \atp concentration in \fig{trajectories}(C) is reversed to forward movement
with $\avg{\vb} = \avg{\veff} \simeq 160 \nmpers > 0$. Although motor
cycle time is increased by the decreased \atp concentration, load
sharing by an increased number of motors leads to larger and eventually
positive positional steps per motor cycle. For sufficiently large load, the
mechanical effect of load sharing outruns the effect of motor cycle
kinetics.


\section{Phase diagrams}

\begin{figure}[t]
\includegraphics[width=0.8\columnwidth]{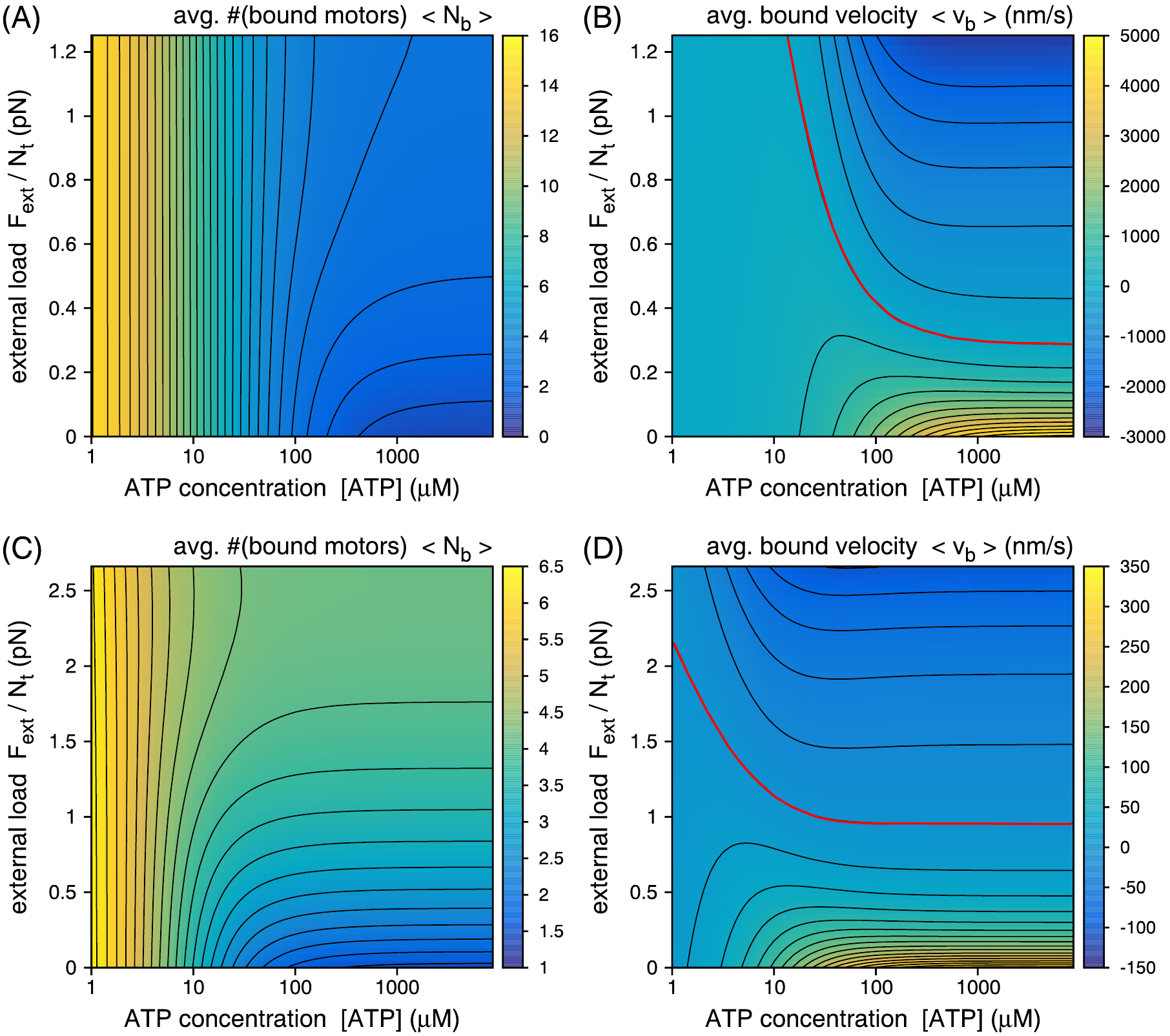}
\caption{Muscle isoforms. Phase diagrams for (A) the average number $\avg{\nb}$ of bound
motors and (B) the average bound velocity $\avg{\vb}$ for a skeletal muscle \myo ensemble of
size $\nt = 16$ as a function of external load per motor $\fext / \nt$
and \atp concentration $\catp$. (C,D) The same for smooth muscle \myo, but for smaller ensemble size $\nt = 8$. The stall force is marked
by a thick red line.}
\label{fig:averages} 
\end{figure}

We now turn to a systematic analysis of the averaged behavior of small \myo
ensembles. \fig{averages}(A) and (B) show average number of bound
motors and average bound velocity, respectively, as function of \atp concentration and
external load for a small ensemble with skeletal muscle \myo.  For small
\atp concentrations, the average number of bound motors shown in \fig{averages}(A) is large and
independent of $\fext$, because the motor cycle is limited by unbinding
from rigor state. With increasing \atp concentration, $\avg{\nb}$
decreases rapidly and becomes load dependent. Above physiological \atp
concentration of $\sim 1\miM$, the motor cycle is limited by load
dependent rates $\kdv + \kdf \ll \kvf$, thus $\avg{\nb}$ becomes
independent of \atp concentration.

\fig{averages}(B) reveals a similar pattern for $\avg{\vb}$ as for
$\avg{\nb}$, with weak load dependence for small \atp concentrations and
weak \atp dependence at high \atp concentrations.  However, the behavior
of $\avg{\vb}$ with increasing \atp concentration is more complex than
for $\avg{\nb}$. For very small $\fext$, $\avg{\vb}$ increases
monotonously with \atp concentration because the motor cycle is
accelerated.  For larger external load, $\avg{\vb}$ increases with small \atp
concentration but passes through a maximum and decreases with further
increasing $\catp$, because the external load is focused on a decreasing
number of bound motors so that the motor cycle leads to backward steps
of the ensemble. For \atp concentrations above the physiological level,
$\avg{\vb}$ becomes independent of $\catp$, but depends strongly
on $\fext$. The upward convex force-velocity relation at constant $\catp$
corresponds to the Hill relation \cite{a:Hill1938} and is due to load
sharing by an increasing number of bound motors \cite{a:Duke1999,
a:ErdmannSchwarz2012}. The average effective velocity $\avg{\veff}$
shows a similar behavior as $\avg{\vb}$ (not shown). Because the boundaries
between the different regimes mainly depend
on single motor properties, they shift only slightly for larger ensemble
size $\nt$ (not shown). The typical level of load which \myo ensembles can sustain is specified
by bound and effective stall forces, $\fbs$ and $\fes$, at which average
bound and effective velocities vanish, respectively. Marked by the red
isoline in \fig{averages}(B), the bound stall force $\fbs$ decreases
strongly with increasing \atp concentration. This implies that to achieve
forward motion, it is better to work at low \atp concentration. Due to stochastic ensemble
detachment, the effective stall force is slightly smaller
than $\fbs$ for $\catp > 100 \muM$.

Having first discussed skeletal \myo as a reference case, we next
turn to smooth muscle \myo. As evident from \tab{general_rates},
the most important change in the parameter set for smooth muscle \myo
relative to skeletal muscle \myo are the small values of the transition
rates $\kd^0$ from post-powerstroke state $(3)$ and the rate $\kez$ of
binding to the weakly-bound state $(2)$. At vanishing load and large
\atp concentration, these rates lead to a single motor duty ratio of
$\rho \simeq 0.25$ compared to $\rho \simeq 0.1$ for skeletal muscle
\myo. Therefore, a significantly smaller ensemble size $\nt$ is sufficient
to stabilize ensemble attachment.

\fig{averages}(C) and (D) shows the average number of bound motors and the
average bound velocity, respectively, of an ensemble of $\nt = 8$ smooth muscle \myo
motors as function of \atp concentration and external load. The plots
reveal the same qualitative dependence of $\avg{\nb}$ and $\avg{\vb}$ on
$\catp$ and $\fext/\nt$ as for skeletal muscle \myo. However, the transition from the \atp sensitive
regime (at low $\catp$) to the load sensitive regime (at large $\catp$)
is shifted to smaller \atp concentrations because of the smaller value
of $\kdv + \kdf$ relative to the rate $\kvf = \katp\catp$ of unbinding
from rigor state at a given value of $\catp$. This effect is partially
offset by the smaller value of the \atp binding rate $\kt$. At small
\atp concentrations, the fraction of bound motors is comparable
(although slightly smaller because of the reduced binding rate $\kez$)
to the case of skeletal muscle \myo. In the load sensitive regime at
large \atp, on the other hand, the average fraction $\avg{\nb}/\nt$ of
bound motors is significantly larger because of the increased single
motor duty ratio.  Moreover, $\avg{\nb}/\nt$ increases more strongly
with $\fext/\nt$ (from $\avg{\nb}/\nt \simeq 0.18$ at $\fext = 0$ to a
maximum of $\avg{\nb}/\nt \simeq 0.52$ at $\fext/\nt \simeq 2.5\pN$),
because the force scale $\fc$ for the catch-path is smaller than for
skeletal muscle \myo. For very small \atp concentrations, bound velocity
is mainly determined by the slow unbinding from rigor state $(4)$ and
becomes comparable for smooth and skeletal muscle \myo. For large \atp
concentrations, bound velocity $\avg{\vb}$ is reduced by a factor $\sim
10$ because of the reduced rates $\kdv + \kdf$ and $\kez$. Due to the
larger fraction of bound motors sharing the external load, however,
$\avg{\vb}$ reduces more slowly with increasing load and the stall force
per bond, $\fbs/\nt$, is larger than for skeletal muscle \myo (note the
larger force scale in \fig{averages}(D) compared to \fig{averages}(B)).

\begin{figure}[t]
\includegraphics[width=0.8\columnwidth]{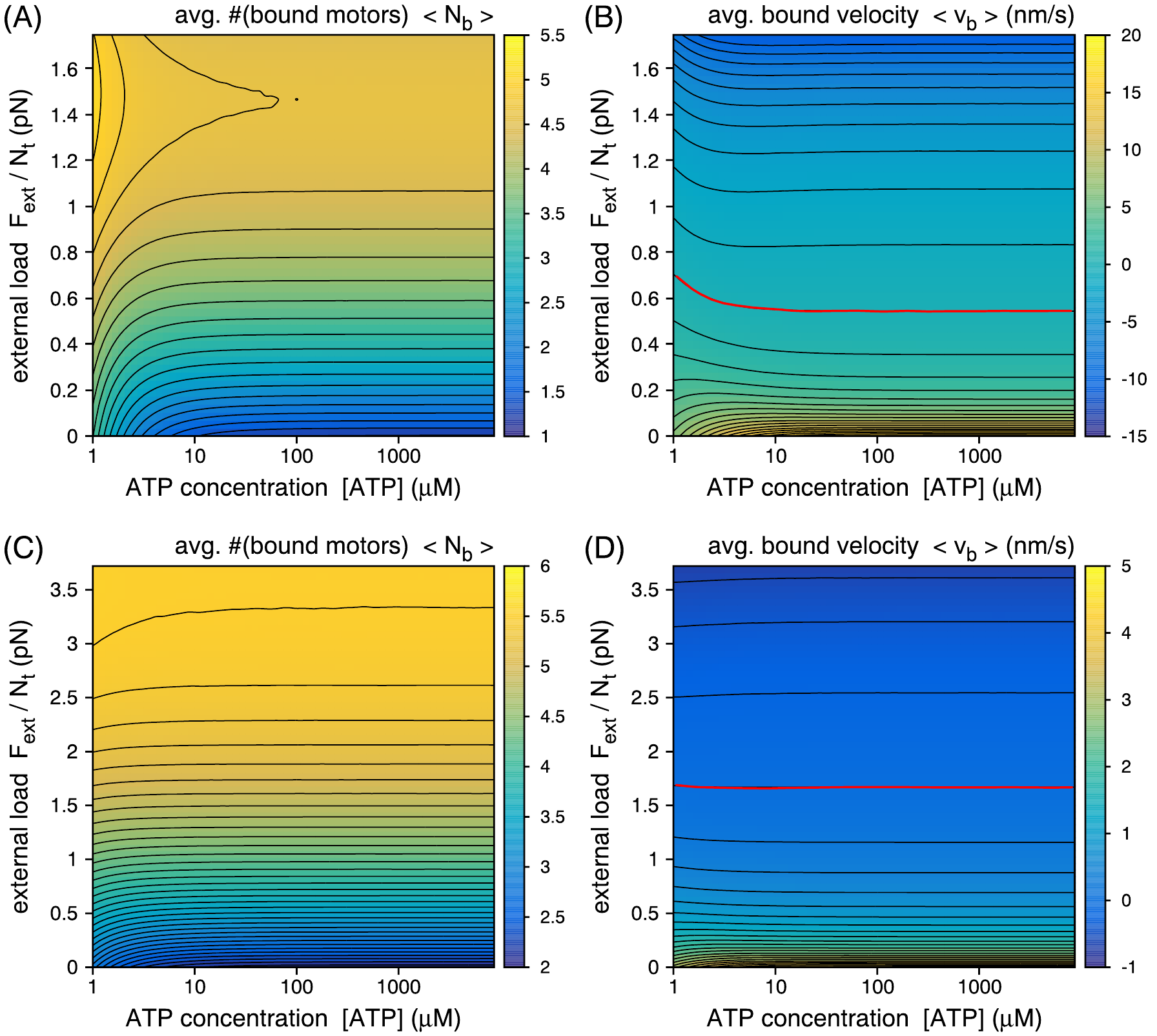}
\caption{Non-muscle isoforms. Phase diagrams for
(A) the average number $\avg{\nb}$ of bound motors and (B) the average
bound velocity $\avg{\vb}$ for a non-muscle \myoA ensemble of size $\nt = 16$ as a
function of external load per motor $\fext / \nt$ and \atp concentration $\catp$.
(C,D) The same for non-muscle \myoB, but for smaller ensemble
size $\nt = 8$. The stall force is marked by a thick red line.}
\label{fig:averages_nm} 
\end{figure}

We next discuss the cases of the non-muscle \myo isoforms. As in the case of skeletal and
smooth muscle \myo, mechanical parameters for the two non-muscle
isoforms of \myo are very similar. Compared to the muscle isoforms,
however, neck linker stiffness $\km$ is larger and powerstroke length
$d$ is smaller for the non-muscle isoforms. Dynamics of non-muscle \myoA
and B is characterized by very small values of binding rate $\kez$ and
transition rate $\kdv + \kdf$, which slow down the motor cycle. The
values of the transition rates result in single motor duty ratios at
vanishing load and large \atp concentration of $\rho \simeq 0.1$ for
non-muscle \myoA and $\rho \simeq 0.36$ for non-muscle \myoB.

\fig{averages_nm}(A) and (B) show load and \atp dependence of $\avg{\nb}$ and
$\avg{\vb}$, respectively, for an ensemble of non-muscle \myoA with $\nt = 16$,
so that it can be compared well with skeletal muscle \myo from \fig{averages}(A) and (B).
In contrast to the cases of the muscle isoforms, $\avg{\nb}$ and $\avg{\vb}$ are
essentially independent of \atp concentration due to the very small
value of the load-dependent rate $\kdv + \kdf$ relative to
the \atp dependent rate $\kvf$. Thus $\avg{\nb}$ and $\avg{\vb}$
are load dependent over the full range of \atp concentrations.
The average number of bound motors is
similar to the case of skeletal muscle \myo in the load dependent
regime at high \atp concentration. The average bound velocity for non-muscle \myoA is very small
compared to the case of skeletal muscle \myo, but shows the same
Hill-type decrease with increasing load. The bound stall force $\fbs$ is
essentially independent of \atp concentration. 

For non-muscle \myoB, transition rate $\kdv + \kdf$ from the
post-powerstroke state as given in \tab{general_rates} is further reduced relative to non-muscle \myoA,
compare \tab{general_rates}. As a consequence, non-muscle \myoB has a higher single motor duty ratio
$\rho \simeq 0.36$, but the motor cycle is even slower than for
non-muscle \myoA. This relation of the non-muscle isoform is similar to
the relation of slow smooth muscle \myo with large duty ratio to the
fast skeletal muscle \myo with a small duty ratio.

\fig{averages_nm}(C) and (D) shows the average number of bound motors and
average bound velocity of an ensemble of non-muscle \myoB
motors as function of \atp concentration and external load. 
Here we choose $\nt = 8$ in order to compare
with the smooth muscle case from \fig{averages}(C) and (D).
The transition to the \atp sensitive regime is shifted to even smaller \atp concentrations
than for non-muscle \myoA. As expected from the
larger single motor duty ratio, the average fraction $\avg{\nb}/\nt$ of
bound motors is larger than for non-muscle \myoA or smooth muscle \myo
in the load sensitive regime at large $\catp$. Due to the smaller value
of $\kdv + \kdf$ the average bound velocity $\avg{\vb}$ is further
reduced.  Because of the large fraction of bound motors and the large
value of neck linker stiffness, however, the bound stall force per motor
is significantly larger than for non-muscle \myoA or smooth muscle
\myo.

\begin{figure}[t]
\includegraphics[width=0.8\columnwidth]{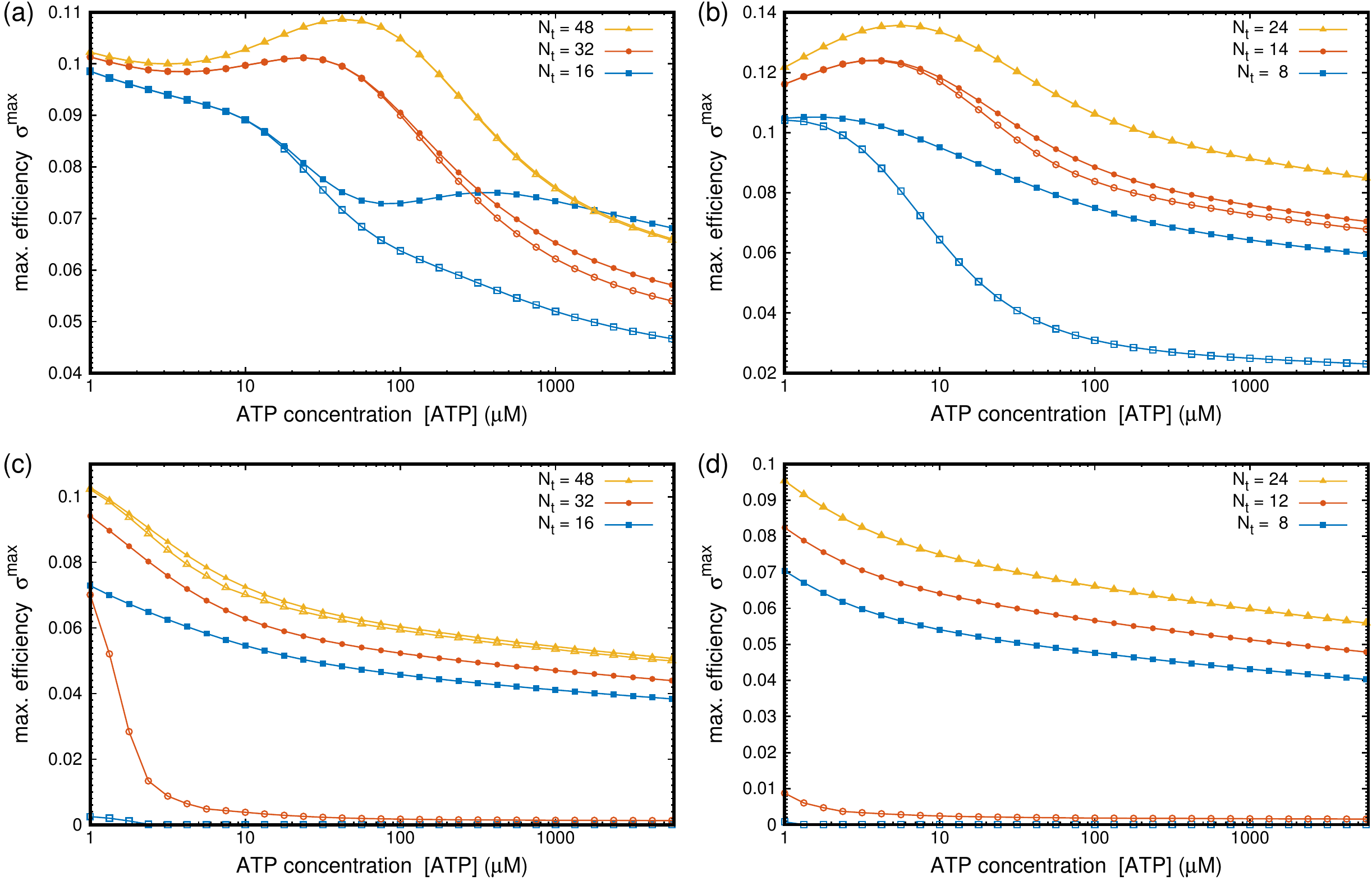}
\caption{Maximal efficiency $\sbm$ and $\sem$ for the bound movement
(solid symbols) and the effective movement (open symbols)
as function of \atp concentration.
(A) Skeletal muscle \myo ensembles with sizes $\nt = 16$, $24$ and $48$. 
(B) Smooth muscle \myo ensembles with sizes $\nt = 8$, $14$ and $24$.
(C) Non-muscle \myoA ensembles with same sizes as in (A).
(D) Non-muscle \myoB ensembles with sizes $\nt = 8$, $12$ and $24$.}
\label{fig:efficiency}
\end{figure}

\section{Ensemble efficiency}

The observation that decreasing \atp
concentration can increase the average bound velocity of a \myo ensemble
means that the efficiency of movement can be increased by a reduced
energy supply. To investigate this interesting point in more detail, we define the effective
thermodynamic efficiency for bound and effective movement as the ratio
of power output and input \cite{a:ParmeggianiEtAl1999,
a:SchmiedlSeifert2008, a:ZimmermannSeifert2012}:
\begin{equation}\label{eq:efficiency}
\sb = \frac{\fext\avg{\vb}}{\jatp\DelG} \quad\text{and}\quad
\seff = \frac{\fext\avg{\veff}}{\jatp\DelG}\,.
\end{equation}
$\jatp$ is the average flux through the motor cycle, in which \atp is
converted to \adp and \pho, and $\DelG$ is the Gibbs free energy
released during \atp hydrolysis. For convenience, we calculate the flux
for the load-independent transition $(1) \to (2)$ as $\jatp = \nt (p_1
k_{12} - p_2 k_{21})$, where $p_i$ is the stationary probability to be
in state $i$, thereby neglecting small corrections that might result
from load dependance. $\DelG$ depends on \atp concentration through
$\DelG = \DelG_0 - \kb T \ln\left(\catp M / \cadp\cpho\right)$ \cite{b:Howard2001},
where concentrations are measured in units of $M$
and $\DelG_0 = -13\ \kb T$, $\cadp = 10 \muM$ and $\cpho = 1\miM$ under
physiological conditions.

\fig{efficiency}(A) shows the maximal efficiencies $\sbm$ and $\sem$ for
bound and effective movement as function of \atp concentration for
different skeletal muscle \myo ensemble sizes.
The larger ensemble size, the smaller are the differences between
bound and effective efficiencies. For small \atp concentrations, $\sbm$ and $\sem$ depend
weakly on $\catp$ because both flux $\jatp$ and power output approach
zero for $\catp \to 0$. For larger $\nt$, $\sbm$ and $\sem$ display a
maximum before decreasing with increasing \atp
concentrations. Above physiological \atp concentrations, $\sbm$ and
$\sbm$ continue to decrease slowly because the maximal power output plateaus
and energy consumption $\jatp\DelG$ continues to increase. Behavior of
$\sbm$ and $\sem$ confirms that reducing energy supply increases
ensemble efficiency, in particular for \atp concentrations just below
the physiological level. 

The maximal efficiencies $\sbm$ and $\sem$ plotted in
\fig{efficiency}(B) for ensembles of smooth muscle \myo show the
same qualitative dependence on $\catp$ as observed for skeletal muscle
\myo. Because the transition to the \atp sensitive regime occurs at
smaller values of the \atp concentration, the maxima of $\sbm$ and
$\sem$ are also shifted to smaller \atp concentrations. For the smallest
ensemble size $\nt = 8$, the difference of effective and bound
efficiencies is larger than for skeletal muscle \myo and is observed at
smaller \atp concentrations. Because of the smaller bound velocity and
the reduced rebinding rate, ensemble detachment reduces the effective
velocity more strongly than for skeletal muscle \myo. Although ensemble
velocity is smaller for smooth muscle \myo, the efficiency is
quantitatively similar to the case of skeletal muscle \myo, because the
reduced power output is compensated by the reduced rate of \atp
consumption.

\fig{efficiency}(C) shows the maximal efficiencies $\sbm$ and $\sem$ for
non-muscle \myoA. For bound movement $\sbm$ shows a significant decrease
with increasing \atp concentration only for extremely small values of
$\catp$. Maximal efficiency $\sem$ of effective movement deviates
strongly from $\sbm$ and drops to zero for the smaller ensembles as a
consequence of ensemble detachment. Although ensemble detachment does
not occur more frequently than for skeletal muscle \myo, effective
velocity is reduced more strongly because of the smaller rebinding rate
$\kez$ and the large unbound velocity relative to $\avg{\vb}$.

\fig{efficiency}(D) plots the maximal efficiencies of bound and
effective ensemble movement for non-muscle \myoB ensembles. The maximal efficiency $\sbm$ of
the bound ensemble movement shows a very weak decrease with increasing
\atp concentration similar to non-muscle \myoA and comparable to
skeletal and smooth muscle \myo at large values of $\catp$. Although
bound ensemble velocity is very small for non-muscle \myoB, the
efficiency is quantitatively similar to the case of skeletal muscle
\myo, because the reduced power output is compensated by the reduced
rate of \atp consumption. As for non-muscle \myoA, the maximal
efficiency $\sem$ of effective movement deviates strongly from $\sbm$
and drops to zero for the smaller ensembles as a consequence of ensemble
detachment. Because of the smaller rebinding rate $\kez$ and the large
unbound velocity $\vu$ relative to $\avg{\vb}$, the effective velocity
becomes negative for very small external load on the ensemble.


\section{Discussion}

Using a detailed five-state crossbridge model for the
\myo motor cycle, we have systematically analyzed the influence of mechanical load and
\atp concentration on the stochastic dynamics of small \myo ensembles
for different isoforms of \myo. Because load and \atp dependence are
described by sequential transitions in the crossbridge cycle, influence
of load becomes more pronounced with increasing \atp concentration.

For the muscle isoforms we
observe two distinct regimes for \myo ensemble dynamics: an \atp
sensitive regime with weak load dependence at small \atp concentrations,
and a mechano-sensitive regime at large \atp concentrations. For
the non-muscle isoforms, which cycle much slower than their
muscle counterparts, only the
mechano-sensitive regime is observed. Transition to an \atp sensitive
regime would require \atp concentrations well below the level commonly
used in motility assays or actomyosin gels. We speculate that 
\atp concentrations in cells might be locally more variable than formerly
appreciated \cite{imamura2009visualization, nakano2011ca2+,
ando2012visualization}, for example during phases of fast actin polymerization and strong 
actomyosin contraction in migrating cells,
but that the non-muscle isoforms are buffered from this effect
by their low \atp sensitivity as demonstrated in \fig{averages_nm}
compared to \fig{averages}.

Ensemble movement results from the interplay of motor cycle kinetics and
ensemble mechanics which are both affected by \atp concentration.
Decreasing \atp concentration from the mechano-sensitive regime at near
vanishing load stabilizes ensembles but decreases velocity. This is
known from skeletal muscle and was investigated before in motility
assays \cite{a:WalcottEtAl2012}. Here we also
have shown the effects for decreasing \atp concentration at large
load, as it might occur in the cytoskeleton of non-muscle cells and in
actomyosin gels, and have found that ensemble velocity
can in fact increase, because the collective
effect of load sharing by an increasing number of bound motors outruns
increased motor cycle time. We find that maximal efficiency
increases with decreasing \atp concentration, similar to ratchet models
for single motors \cite{a:ParmeggianiEtAl1999}. For
the small \myo ensembles, however, we find that in our model the
effective thermodynamic efficiency is rather low (typically
$\sbm$ is below $0.1$).

Our results for ensemble efficiency are in stark contrast to the much higher values for single motors,
like e.g.\ the F$_1$-ATPase \cite{a:ZimmermannSeifert2012}. They are also in stark contrast
to the values for skeletal muscle, which has been measured to be
of the order of $0.3$ \cite{whipp1969efficiency}. There are several reasons why
efficiency is low in our model. We first note that motors mechanically
work against each other and that they dissipate elastic energy during unbinding.
We also note that for small ensembles, our results are strongly shaped by the physical process
that takes over during times of unbinding. For simplicity, here we have used hydrodynamic
slip during times of unbinding, but it would be interesting to consider also other physical processes
in this context. Interestingly, we also observed that in our model, efficiency can be as high as $0.5$ when
optimizing parameter values (mainly by increasing $\km$ while
keeping \atp flux effectively constant by adjusting other parameters).
This indicates that our results depend sensitively
on parameter values, which here have been chosen from the literature
as listed in \tab{general_rates}.

Finally, our work shows that one has to be careful when drawing
conclusions on cellular contractility from reconstituted actomyosin
gels. Here one often uses skeletal or smooth muscle \myo 
and reduces \atp concentrations to stabilize the system \cite{a:SoaresEtAl2011,kohler2011structure,
murrell2012f,ideses2013myosin,linsmeier_disordered_2016}.
Our analysis shows that decreasing \atp concentrations
has the desired effect of increased contractility for the muscle \myo isoforms. However,
it also shows that the same would not work for non-muscle \myo isoforms,
because they are less sensitive to changes in \atp concentrations,
and that the resulting numbers for bound motors and contraction velocities
might be quite different.

\begin{acknowledgments}
USS is a member of the Interdisciplinary Center for Scientific Computing
(IWR) and of the cluster of excellence CellNetworks at Heidelberg. We
thank Sam Walcott for helpful discussions and Philipp Albert and Marcel Weiss
for critical reading of the manuscript.
\end{acknowledgments}


\begin{thebibliography}{38}
\expandafter\ifx\csname natexlab\endcsname\relax\def\natexlab#1{#1}\fi
\expandafter\ifx\csname bibnamefont\endcsname\relax
  \def\bibnamefont#1{#1}\fi
\expandafter\ifx\csname bibfnamefont\endcsname\relax
  \def\bibfnamefont#1{#1}\fi
\expandafter\ifx\csname citenamefont\endcsname\relax
  \def\citenamefont#1{#1}\fi
\expandafter\ifx\csname url\endcsname\relax
  \def\url#1{\texttt{#1}}\fi
\expandafter\ifx\csname urlprefix\endcsname\relax\def\urlprefix{URL }\fi
\providecommand{\bibinfo}[2]{#2}
\providecommand{\eprint}[2][]{\url{#2}}

\bibitem[{\citenamefont{Howard}(2001)}]{b:Howard2001}
\bibinfo{author}{\bibfnamefont{J.}~\bibnamefont{Howard}},
  \emph{\bibinfo{title}{{Mechanics of Motor Proteins and the Cytoskeleton}}}
  (\bibinfo{publisher}{Sinauer Associates}, \bibinfo{address}{Sunderland, MA},
  \bibinfo{year}{2001}).

\bibitem[{\citenamefont{Vicente-Manzanares
  et~al.}(2009)\citenamefont{Vicente-Manzanares, Ma, Adelstein, and
  Horwitz}}]{rv:VicenteEtAl2009}
\bibinfo{author}{\bibfnamefont{M.}~\bibnamefont{Vicente-Manzanares}},
  \bibinfo{author}{\bibfnamefont{X.}~\bibnamefont{Ma}},
  \bibinfo{author}{\bibfnamefont{R.~S.} \bibnamefont{Adelstein}},
  \bibnamefont{and} \bibinfo{author}{\bibfnamefont{A.~R.}
  \bibnamefont{Horwitz}}, \bibinfo{journal}{Nat. Rev. Mol. Cell Biol.}
  \textbf{\bibinfo{volume}{10}}, \bibinfo{pages}{778} (\bibinfo{year}{2009}).

\bibitem[{\citenamefont{Murrell et~al.}(2015)\citenamefont{Murrell, Oakes,
  Lenz, and Gardel}}]{murrell_forcing_2015}
\bibinfo{author}{\bibfnamefont{M.}~\bibnamefont{Murrell}},
  \bibinfo{author}{\bibfnamefont{P.~W.} \bibnamefont{Oakes}},
  \bibinfo{author}{\bibfnamefont{M.}~\bibnamefont{Lenz}}, \bibnamefont{and}
  \bibinfo{author}{\bibfnamefont{M.~L.} \bibnamefont{Gardel}},
  \bibinfo{journal}{Nature Reviews. Molecular Cell Biology}
  \textbf{\bibinfo{volume}{16}}, \bibinfo{pages}{486} (\bibinfo{year}{2015}).

\bibitem[{\citenamefont{Engler et~al.}(2006)\citenamefont{Engler, Sen, Sweeney,
  and Discher}}]{a:EnglerEtAl2006}
\bibinfo{author}{\bibfnamefont{A.~J.} \bibnamefont{Engler}},
  \bibinfo{author}{\bibfnamefont{S.}~\bibnamefont{Sen}},
  \bibinfo{author}{\bibfnamefont{H.~L.} \bibnamefont{Sweeney}},
  \bibnamefont{and} \bibinfo{author}{\bibfnamefont{D.~E.}
  \bibnamefont{Discher}}, \bibinfo{journal}{Cell}
  \textbf{\bibinfo{volume}{126}}, \bibinfo{pages}{677} (\bibinfo{year}{2006}).

\bibitem[{\citenamefont{Billington et~al.}(2013)\citenamefont{Billington, Wang,
  Mao, Adelstein, and Sellers}}]{billington_characterization_2013}
\bibinfo{author}{\bibfnamefont{N.}~\bibnamefont{Billington}},
  \bibinfo{author}{\bibfnamefont{A.}~\bibnamefont{Wang}},
  \bibinfo{author}{\bibfnamefont{J.}~\bibnamefont{Mao}},
  \bibinfo{author}{\bibfnamefont{R.~S.} \bibnamefont{Adelstein}},
  \bibnamefont{and} \bibinfo{author}{\bibfnamefont{J.~R.}
  \bibnamefont{Sellers}}, \bibinfo{journal}{Journal of Biological Chemistry}
  \textbf{\bibinfo{volume}{288}}, \bibinfo{pages}{33398}
  (\bibinfo{year}{2013}).

\bibitem[{\citenamefont{Beach and Hammer~III}(2015)}]{rv:BeachHammer2015}
\bibinfo{author}{\bibfnamefont{J.~R.} \bibnamefont{Beach}} \bibnamefont{and}
  \bibinfo{author}{\bibfnamefont{J.~A.} \bibnamefont{Hammer~III}},
  \bibinfo{journal}{Exp. Cell Res.} \textbf{\bibinfo{volume}{334}},
  \bibinfo{pages}{2} (\bibinfo{year}{2015}).

\bibitem[{\citenamefont{Erdmann and Schwarz}(2012)}]{a:ErdmannSchwarz2012}
\bibinfo{author}{\bibfnamefont{T.}~\bibnamefont{Erdmann}} \bibnamefont{and}
  \bibinfo{author}{\bibfnamefont{U.~S.} \bibnamefont{Schwarz}},
  \bibinfo{journal}{Phys. Rev. Lett.} \textbf{\bibinfo{volume}{108}},
  \bibinfo{pages}{188101} (\bibinfo{year}{2012}).

\bibitem[{\citenamefont{Veigel and Schmidt}(2011)}]{rv:VeigelSchmidt2011}
\bibinfo{author}{\bibfnamefont{C.}~\bibnamefont{Veigel}} \bibnamefont{and}
  \bibinfo{author}{\bibfnamefont{C.~F.} \bibnamefont{Schmidt}},
  \bibinfo{journal}{Nat. Rev. Mol. Cell Biol.} \textbf{\bibinfo{volume}{12}},
  \bibinfo{pages}{163} (\bibinfo{year}{2011}).

\bibitem[{\citenamefont{Walcott et~al.}(2012)\citenamefont{Walcott, Warshaw,
  and Debold}}]{a:WalcottEtAl2012}
\bibinfo{author}{\bibfnamefont{S.}~\bibnamefont{Walcott}},
  \bibinfo{author}{\bibfnamefont{D.~M.} \bibnamefont{Warshaw}},
  \bibnamefont{and} \bibinfo{author}{\bibfnamefont{E.~P.}
  \bibnamefont{Debold}}, \bibinfo{journal}{Biophys. J.}
  \textbf{\bibinfo{volume}{103}}, \bibinfo{pages}{501} (\bibinfo{year}{2012}).

\bibitem[{\citenamefont{Hilbert et~al.}(2013)\citenamefont{Hilbert, Cumarasamy,
  Zitouni, Mackey, and Lauzon}}]{a:HilbertEtAl2013}
\bibinfo{author}{\bibfnamefont{L.}~\bibnamefont{Hilbert}},
  \bibinfo{author}{\bibfnamefont{S.}~\bibnamefont{Cumarasamy}},
  \bibinfo{author}{\bibfnamefont{N.~B.} \bibnamefont{Zitouni}},
  \bibinfo{author}{\bibfnamefont{M.~C.} \bibnamefont{Mackey}},
  \bibnamefont{and} \bibinfo{author}{\bibfnamefont{A.-M.}
  \bibnamefont{Lauzon}}, \bibinfo{journal}{Biophys. J.}
  \textbf{\bibinfo{volume}{105}}, \bibinfo{pages}{1466} (\bibinfo{year}{2013}).

\bibitem[{\citenamefont{Soares~e Silva et~al.}(2011)\citenamefont{Soares~e
  Silva, Depken, Stuhrmann, Korsten, MacKintosh, and
  Koenderink}}]{a:SoaresEtAl2011}
\bibinfo{author}{\bibfnamefont{M.}~\bibnamefont{Soares~e Silva}},
  \bibinfo{author}{\bibfnamefont{M.}~\bibnamefont{Depken}},
  \bibinfo{author}{\bibfnamefont{B.}~\bibnamefont{Stuhrmann}},
  \bibinfo{author}{\bibfnamefont{M.}~\bibnamefont{Korsten}},
  \bibinfo{author}{\bibfnamefont{F.~C.} \bibnamefont{MacKintosh}},
  \bibnamefont{and} \bibinfo{author}{\bibfnamefont{G.~H.}
  \bibnamefont{Koenderink}}, \bibinfo{journal}{Proc. Natl. Acad. Sci. USA}
  \textbf{\bibinfo{volume}{108}}, \bibinfo{pages}{9408} (\bibinfo{year}{2011}).

\bibitem[{\citenamefont{K{\"o}hler et~al.}(2011)\citenamefont{K{\"o}hler,
  Schaller, and Bausch}}]{kohler2011structure}
\bibinfo{author}{\bibfnamefont{S.}~\bibnamefont{K{\"o}hler}},
  \bibinfo{author}{\bibfnamefont{V.}~\bibnamefont{Schaller}}, \bibnamefont{and}
  \bibinfo{author}{\bibfnamefont{A.~R.} \bibnamefont{Bausch}},
  \bibinfo{journal}{Nature materials} \textbf{\bibinfo{volume}{10}},
  \bibinfo{pages}{462} (\bibinfo{year}{2011}).

\bibitem[{\citenamefont{Murrell and Gardel}(2012)}]{murrell2012f}
\bibinfo{author}{\bibfnamefont{M.~P.} \bibnamefont{Murrell}} \bibnamefont{and}
  \bibinfo{author}{\bibfnamefont{M.~L.} \bibnamefont{Gardel}},
  \bibinfo{journal}{Proceedings of the National Academy of Sciences}
  \textbf{\bibinfo{volume}{109}}, \bibinfo{pages}{20820}
  (\bibinfo{year}{2012}).

\bibitem[{\citenamefont{Ideses et~al.}(2013)\citenamefont{Ideses, Sonn-Segev,
  Roichman, and Bernheim-Groswasser}}]{ideses2013myosin}
\bibinfo{author}{\bibfnamefont{Y.}~\bibnamefont{Ideses}},
  \bibinfo{author}{\bibfnamefont{A.}~\bibnamefont{Sonn-Segev}},
  \bibinfo{author}{\bibfnamefont{Y.}~\bibnamefont{Roichman}}, \bibnamefont{and}
  \bibinfo{author}{\bibfnamefont{A.}~\bibnamefont{Bernheim-Groswasser}},
  \bibinfo{journal}{Soft Matter} \textbf{\bibinfo{volume}{9}},
  \bibinfo{pages}{7127} (\bibinfo{year}{2013}).

\bibitem[{\citenamefont{Linsmeier et~al.}(2016)\citenamefont{Linsmeier,
  Banerjee, Oakes, Jung, Kim, and Murrell}}]{linsmeier_disordered_2016}
\bibinfo{author}{\bibfnamefont{I.}~\bibnamefont{Linsmeier}},
  \bibinfo{author}{\bibfnamefont{S.}~\bibnamefont{Banerjee}},
  \bibinfo{author}{\bibfnamefont{P.~W.} \bibnamefont{Oakes}},
  \bibinfo{author}{\bibfnamefont{W.}~\bibnamefont{Jung}},
  \bibinfo{author}{\bibfnamefont{T.}~\bibnamefont{Kim}}, \bibnamefont{and}
  \bibinfo{author}{\bibfnamefont{M.~P.} \bibnamefont{Murrell}},
  \bibinfo{journal}{Nature Communications} \textbf{\bibinfo{volume}{7}},
  \bibinfo{pages}{12615} (\bibinfo{year}{2016}).

\bibitem[{\citenamefont{Josephs and Harrington}(1966)}]{josephs1966studies}
\bibinfo{author}{\bibfnamefont{R.}~\bibnamefont{Josephs}} \bibnamefont{and}
  \bibinfo{author}{\bibfnamefont{W.~F.} \bibnamefont{Harrington}},
  \bibinfo{journal}{Biochemistry} \textbf{\bibinfo{volume}{5}},
  \bibinfo{pages}{3474} (\bibinfo{year}{1966}).

\bibitem[{\citenamefont{Niederman and Pollard}(1975)}]{niederman1975human}
\bibinfo{author}{\bibfnamefont{R.}~\bibnamefont{Niederman}} \bibnamefont{and}
  \bibinfo{author}{\bibfnamefont{T.~D.} \bibnamefont{Pollard}},
  \bibinfo{journal}{The Journal of Cell Biology} \textbf{\bibinfo{volume}{67}},
  \bibinfo{pages}{72} (\bibinfo{year}{1975}).

\bibitem[{\citenamefont{Somlyo and Somlyo}(2003)}]{somlyo2003ca2+}
\bibinfo{author}{\bibfnamefont{A.~P.} \bibnamefont{Somlyo}} \bibnamefont{and}
  \bibinfo{author}{\bibfnamefont{A.~V.} \bibnamefont{Somlyo}},
  \bibinfo{journal}{Physiological reviews} \textbf{\bibinfo{volume}{83}},
  \bibinfo{pages}{1325} (\bibinfo{year}{2003}).

\bibitem[{\citenamefont{Piazzesi et~al.}(2007)\citenamefont{Piazzesi,
  Reconditi, Linari, Lucii, Bianco, Brunello, Decostre, Stewart, Gore, Irving
  et~al.}}]{a:PiazzesiEtAl2007}
\bibinfo{author}{\bibfnamefont{G.}~\bibnamefont{Piazzesi}},
  \bibinfo{author}{\bibfnamefont{M.}~\bibnamefont{Reconditi}},
  \bibinfo{author}{\bibfnamefont{M.}~\bibnamefont{Linari}},
  \bibinfo{author}{\bibfnamefont{L.}~\bibnamefont{Lucii}},
  \bibinfo{author}{\bibfnamefont{P.}~\bibnamefont{Bianco}},
  \bibinfo{author}{\bibfnamefont{E.}~\bibnamefont{Brunello}},
  \bibinfo{author}{\bibfnamefont{V.}~\bibnamefont{Decostre}},
  \bibinfo{author}{\bibfnamefont{A.}~\bibnamefont{Stewart}},
  \bibinfo{author}{\bibfnamefont{D.~B.} \bibnamefont{Gore}},
  \bibinfo{author}{\bibfnamefont{T.~C.} \bibnamefont{Irving}},
  \bibnamefont{et~al.}, \bibinfo{journal}{Cell} \textbf{\bibinfo{volume}{131}},
  \bibinfo{pages}{784} (\bibinfo{year}{2007}).

\bibitem[{\citenamefont{Veigel et~al.}(2003)\citenamefont{Veigel, Molloy,
  Schmitz, and Kendrick-Jones}}]{a:VeigelEtAl2003}
\bibinfo{author}{\bibfnamefont{C.}~\bibnamefont{Veigel}},
  \bibinfo{author}{\bibfnamefont{J.~E.} \bibnamefont{Molloy}},
  \bibinfo{author}{\bibfnamefont{S.}~\bibnamefont{Schmitz}}, \bibnamefont{and}
  \bibinfo{author}{\bibfnamefont{J.}~\bibnamefont{Kendrick-Jones}},
  \bibinfo{journal}{Nat. Cell Biol.} \textbf{\bibinfo{volume}{5}},
  \bibinfo{pages}{980} (\bibinfo{year}{2003}).

\bibitem[{\citenamefont{Albert et~al.}(2014)\citenamefont{Albert, Erdmann, and
  Schwarz}}]{a:AlbertEtAl2014}
\bibinfo{author}{\bibfnamefont{P.~J.} \bibnamefont{Albert}},
  \bibinfo{author}{\bibfnamefont{T.}~\bibnamefont{Erdmann}}, \bibnamefont{and}
  \bibinfo{author}{\bibfnamefont{U.~S.} \bibnamefont{Schwarz}},
  \bibinfo{journal}{New J. Phys.} \textbf{\bibinfo{volume}{16}},
  \bibinfo{pages}{093019} (\bibinfo{year}{2014}).

\bibitem[{\citenamefont{Stam et~al.}(2015)\citenamefont{Stam, Alberts, Gardel,
  and Munro}}]{a:StamEtAl2015}
\bibinfo{author}{\bibfnamefont{S.}~\bibnamefont{Stam}},
  \bibinfo{author}{\bibfnamefont{J.}~\bibnamefont{Alberts}},
  \bibinfo{author}{\bibfnamefont{M.~L.} \bibnamefont{Gardel}},
  \bibnamefont{and} \bibinfo{author}{\bibfnamefont{E.}~\bibnamefont{Munro}},
  \bibinfo{journal}{Biophys. J.} \textbf{\bibinfo{volume}{108}},
  \bibinfo{pages}{1997} (\bibinfo{year}{2015}).

\bibitem[{\citenamefont{Cooke and Bialek}(1979)}]{cooke1979contraction}
\bibinfo{author}{\bibfnamefont{R.}~\bibnamefont{Cooke}} \bibnamefont{and}
  \bibinfo{author}{\bibfnamefont{W.}~\bibnamefont{Bialek}},
  \bibinfo{journal}{Biophysical journal} \textbf{\bibinfo{volume}{28}},
  \bibinfo{pages}{241} (\bibinfo{year}{1979}).

\bibitem[{\citenamefont{Imamura et~al.}(2009)\citenamefont{Imamura, Nhat,
  Togawa, Saito, Iino, Kato-Yamada, Nagai, and
  Noji}}]{imamura2009visualization}
\bibinfo{author}{\bibfnamefont{H.}~\bibnamefont{Imamura}},
  \bibinfo{author}{\bibfnamefont{K.~P.~H.} \bibnamefont{Nhat}},
  \bibinfo{author}{\bibfnamefont{H.}~\bibnamefont{Togawa}},
  \bibinfo{author}{\bibfnamefont{K.}~\bibnamefont{Saito}},
  \bibinfo{author}{\bibfnamefont{R.}~\bibnamefont{Iino}},
  \bibinfo{author}{\bibfnamefont{Y.}~\bibnamefont{Kato-Yamada}},
  \bibinfo{author}{\bibfnamefont{T.}~\bibnamefont{Nagai}}, \bibnamefont{and}
  \bibinfo{author}{\bibfnamefont{H.}~\bibnamefont{Noji}},
  \bibinfo{journal}{Proceedings of the National Academy of Sciences}
  \textbf{\bibinfo{volume}{106}}, \bibinfo{pages}{15651}
  (\bibinfo{year}{2009}).

\bibitem[{\citenamefont{Nakano et~al.}(2011)\citenamefont{Nakano, Imamura,
  Nagai, and Noji}}]{nakano2011ca2+}
\bibinfo{author}{\bibfnamefont{M.}~\bibnamefont{Nakano}},
  \bibinfo{author}{\bibfnamefont{H.}~\bibnamefont{Imamura}},
  \bibinfo{author}{\bibfnamefont{T.}~\bibnamefont{Nagai}}, \bibnamefont{and}
  \bibinfo{author}{\bibfnamefont{H.}~\bibnamefont{Noji}}, \bibinfo{journal}{ACS
  chemical biology} \textbf{\bibinfo{volume}{6}}, \bibinfo{pages}{709}
  (\bibinfo{year}{2011}).

\bibitem[{\citenamefont{Ando et~al.}(2012)\citenamefont{Ando, Imamura, Suzuki,
  Aizaki, Watanabe, Wakita, and Suzuki}}]{ando2012visualization}
\bibinfo{author}{\bibfnamefont{T.}~\bibnamefont{Ando}},
  \bibinfo{author}{\bibfnamefont{H.}~\bibnamefont{Imamura}},
  \bibinfo{author}{\bibfnamefont{R.}~\bibnamefont{Suzuki}},
  \bibinfo{author}{\bibfnamefont{H.}~\bibnamefont{Aizaki}},
  \bibinfo{author}{\bibfnamefont{T.}~\bibnamefont{Watanabe}},
  \bibinfo{author}{\bibfnamefont{T.}~\bibnamefont{Wakita}}, \bibnamefont{and}
  \bibinfo{author}{\bibfnamefont{T.}~\bibnamefont{Suzuki}},
  \bibinfo{journal}{PLoS Pathog} \textbf{\bibinfo{volume}{8}},
  \bibinfo{pages}{e1002561} (\bibinfo{year}{2012}).

\bibitem[{\citenamefont{Duke}(1999)}]{a:Duke1999}
\bibinfo{author}{\bibfnamefont{T.~A.~J.} \bibnamefont{Duke}},
  \bibinfo{journal}{Proc. Natl. Acad. Sci. USA} \textbf{\bibinfo{volume}{96}},
  \bibinfo{pages}{2770} (\bibinfo{year}{1999}).

\bibitem[{\citenamefont{Duke}(2000)}]{a:Duke2000}
\bibinfo{author}{\bibfnamefont{T.~A.~J.} \bibnamefont{Duke}},
  \bibinfo{journal}{Phil. Trans. Roy. Soc. B} \textbf{\bibinfo{volume}{355}},
  \bibinfo{pages}{529} (\bibinfo{year}{2000}).

\bibitem[{\citenamefont{Vilfan and Duke}(2003)}]{a:VilfanDuke2003b}
\bibinfo{author}{\bibfnamefont{A.}~\bibnamefont{Vilfan}} \bibnamefont{and}
  \bibinfo{author}{\bibfnamefont{T.~A.~J.} \bibnamefont{Duke}},
  \bibinfo{journal}{Biophys. J.} \textbf{\bibinfo{volume}{85}},
  \bibinfo{pages}{818} (\bibinfo{year}{2003}).

\bibitem[{\citenamefont{Erdmann et~al.}(2013)\citenamefont{Erdmann, Albert, and
  Schwarz}}]{a:ErdmannEtAl2013}
\bibinfo{author}{\bibfnamefont{T.}~\bibnamefont{Erdmann}},
  \bibinfo{author}{\bibfnamefont{P.~J.} \bibnamefont{Albert}},
  \bibnamefont{and} \bibinfo{author}{\bibfnamefont{U.~S.}
  \bibnamefont{Schwarz}}, \bibinfo{journal}{J. Chem. Phys.}
  \textbf{\bibinfo{volume}{139}}, \bibinfo{pages}{175104}
  (\bibinfo{year}{2013}).

\bibitem[{\citenamefont{Lan and Sun}(2005)}]{a:LanSun2005}
\bibinfo{author}{\bibfnamefont{G.}~\bibnamefont{Lan}} \bibnamefont{and}
  \bibinfo{author}{\bibfnamefont{S.~X.} \bibnamefont{Sun}},
  \bibinfo{journal}{Biophysical Journal} \textbf{\bibinfo{volume}{88}},
  \bibinfo{pages}{4107} (\bibinfo{year}{2005}).

\bibitem[{\citenamefont{Guo and Guilford}(2006)}]{a:GuoGuilford2006}
\bibinfo{author}{\bibfnamefont{B.}~\bibnamefont{Guo}} \bibnamefont{and}
  \bibinfo{author}{\bibfnamefont{W.~H.} \bibnamefont{Guilford}},
  \bibinfo{journal}{Proc. Natl. Acad. Sci. USA} \textbf{\bibinfo{volume}{103}},
  \bibinfo{pages}{9844} (\bibinfo{year}{2006}).

\bibitem[{\citenamefont{Kov{\'a}cs et~al.}(2003)\citenamefont{Kov{\'a}cs, Wang,
  Hu, Zhang, and Sellers}}]{kovacs2003}
\bibinfo{author}{\bibfnamefont{M.}~\bibnamefont{Kov{\'a}cs}},
  \bibinfo{author}{\bibfnamefont{F.}~\bibnamefont{Wang}},
  \bibinfo{author}{\bibfnamefont{A.}~\bibnamefont{Hu}},
  \bibinfo{author}{\bibfnamefont{Y.}~\bibnamefont{Zhang}}, \bibnamefont{and}
  \bibinfo{author}{\bibfnamefont{J.~R.} \bibnamefont{Sellers}},
  \bibinfo{journal}{Journal of Biological Chemistry}
  \textbf{\bibinfo{volume}{278}}, \bibinfo{pages}{38132}
  (\bibinfo{year}{2003}).

\bibitem[{\citenamefont{Hill}(1938)}]{a:Hill1938}
\bibinfo{author}{\bibfnamefont{A.~V.} \bibnamefont{Hill}},
  \bibinfo{journal}{Proc. R. Soc. Lond. B} \textbf{\bibinfo{volume}{126}},
  \bibinfo{pages}{136} (\bibinfo{year}{1938}).

\bibitem[{\citenamefont{Parmeggiani et~al.}(1999)\citenamefont{Parmeggiani,
  J{\"u}licher, Ajdari, and Prost}}]{a:ParmeggianiEtAl1999}
\bibinfo{author}{\bibfnamefont{A.}~\bibnamefont{Parmeggiani}},
  \bibinfo{author}{\bibfnamefont{F.}~\bibnamefont{J{\"u}licher}},
  \bibinfo{author}{\bibfnamefont{A.}~\bibnamefont{Ajdari}}, \bibnamefont{and}
  \bibinfo{author}{\bibfnamefont{J.}~\bibnamefont{Prost}},
  \bibinfo{journal}{Phys. Rev. E} \textbf{\bibinfo{volume}{60}},
  \bibinfo{pages}{2127} (\bibinfo{year}{1999}).

\bibitem[{\citenamefont{Schmiedl and Seifert}(2008)}]{a:SchmiedlSeifert2008}
\bibinfo{author}{\bibfnamefont{T.}~\bibnamefont{Schmiedl}} \bibnamefont{and}
  \bibinfo{author}{\bibfnamefont{U.}~\bibnamefont{Seifert}},
  \bibinfo{journal}{EPL (Europhysics Letters)} \textbf{\bibinfo{volume}{83}},
  \bibinfo{pages}{30005} (\bibinfo{year}{2008}).

\bibitem[{\citenamefont{Zimmermann and
  Seifert}(2012)}]{a:ZimmermannSeifert2012}
\bibinfo{author}{\bibfnamefont{E.}~\bibnamefont{Zimmermann}} \bibnamefont{and}
  \bibinfo{author}{\bibfnamefont{U.}~\bibnamefont{Seifert}},
  \bibinfo{journal}{New J. Phys.} \textbf{\bibinfo{volume}{14}},
  \bibinfo{pages}{103023} (\bibinfo{year}{2012}).

\bibitem[{\citenamefont{Whipp and Wasserman}(1969)}]{whipp1969efficiency}
\bibinfo{author}{\bibfnamefont{B.~J.} \bibnamefont{Whipp}} \bibnamefont{and}
  \bibinfo{author}{\bibfnamefont{K.}~\bibnamefont{Wasserman}},
  \bibinfo{journal}{Journal of Applied Physiology}
  \textbf{\bibinfo{volume}{26}}, \bibinfo{pages}{644} (\bibinfo{year}{1969}).

\end{thebibliography}

\end{document}